\documentstyle[ijmpal]{article}
\def\qed{\hbox{${\vcenter{\vbox{			%HOLLOW SQUARE
   \hrule height 0.4pt\hbox{\vrule width 0.4pt height 6pt
   \kern5pt\vrule width 0.4pt}\hrule height 0.4pt}}}$}}

\def\renewcommand{\thefootnote}{\fnsymbol{footnote}}	%USE SYMBOLIC FOOTNOTE

\begin{document}
 
\runninghead{Coasting Cosmologies
 $\ldots$} {Coasting Cosmologies 
 $\ldots$}
\normalsize\textlineskip
\thispagestyle{empty}
\setcounter{page}{1}

\copyrightheading{}			%{Vol. 0, No. 0 (1993) 000--000}

\vspace*{0.88truein}

\fpage{1}
\centerline{\bf COASTING COSMOLOGIES WITH }
\vspace*{0.37truein}
\centerline{\bf  TIME DEPENDENT COSMOLOGICAL CONSTANT }
\vspace*{0.37truein}

%MANUSCRIPTS USING COMPUTER SOFTWARE
\centerline{\footnotesize LUIS O. PIMENTEL{\footnote{E-mail: lopr@xanum.uam.mx}} \, 
 and LUZ M. DIAZ-RIVERA}
\vspace*{0.015truein}
\centerline{\footnotesize\it Departamento de F{\'\i}sica, Universidad Aut\'onoma 
Metropolitana-Iztapalapa,}
\baselineskip=10pt
\centerline{\footnotesize\it  A. P. 55-534, CP 09340, M\'exico D. F. ,M\'exico}
\vspace*{10pt}
\baselineskip=10pt
\vspace*{0.015truein}
%\centerline{\it and}
%\vspace*{0.015truein}
\baselineskip=10pt
%\centerline{\footnotesize\it City, State ZIP/Zone, Country}
\vspace*{0.225truein}
\publisher{(28.IV.1998)}{(16.VI.1998)}

\vspace*{0.21truein}
\abstracts{
The effect of a time dependent cosmological constant is
considered in a family of scalar-tensor theories.
Friedmann-~Robertson-~Walker cosmological models for vacuum and
perfect fluid matter are found. They have a linear expansion factor, 
the so called coasting cosmology, the gravitational "constant" decreases inversely
with time; that is these models satisfy the Dirac Hypotheses. The
cosmological "constant" decreases inversely with the square of time,
therefore we can have a very small value for it at present time.
}{}{}

%\textlineskip			%) USE THIS MEASUREMENT WHEN THERE IS
%\vspace*{12pt}			%) NO SECTION HEADING

\vspace*{1pt}\textlineskip	%) USE THIS MEASUREMENT WHEN THERE IS
\section{Introduction}	%) A SECTION HEADING
\vspace*{-0.5pt}
\noindent

The renewed interest in the scalar-~tensor theories of gravitation
is caused by two main factors: First, most of the unified theories,
 including super-string theories contain a scalar field (dilaton,
size of the extra compact space in Kaluza-Klein theories) which
play a similar role to the scalar field of the scalar-tensor
theories. Secondly, the new scenario of extended inflation which
solves the fine tuning problem of the old, new and chaotic
inflation has a scalar field that slows the expansion rate of the
universe, from exponential to polynomial, allowing the
completion of the phase transition from the de Sitter phase to a
radiation dominated universe, the graceful exit problem.

In this work we want to consider a family of scalar-~tensor
theories with a potential that is equivalent to a time dependent
cosmological constant. Recently several authors $^{1-12} $  have 
considered the cosmological consequences of 
a time varying cosmological constant. Most of them introduce the time 
dependence in an ad hoc manner. In this work we consider an equivalent 
problem in a the well known general scalar-tensor
 theory of gravity where the time dependence can occur in a natural way,
without any new assumption or modification of the theory.

\section{ Field Equations}
\bigskip

We start our discussion with the action for the most general
scalar-tensor theory of gravitation $^{13}$

\begin{equation}
S={1\over{16\pi G}}\int{d^4x\sqrt{-g}\,[\phi R-\phi^{-1}\omega g^{\mu\nu}
\partial_{\mu}\phi\partial_{\nu}\phi +2 \phi \lambda (\phi )]}+S_{NG},
\end{equation}

\noindent where $g=$ det ( $g_{\mu\nu} $) G is Newton's constant,   
$S_{NG}\quad$ is the action for the nongravitational matter. We shall  
use the signature  $(-,+,+,+)$.
The arbitrary functions $\omega (\phi )$ and
$\lambda (\phi )$ distinguish the different scalar-tensor theories of
 gravitation;
 $\lambda (\phi )$ is a potential function and plays the role of a  
cosmological
 constant, $\omega(\phi) $ is the coupling function
  of the particular theory. General relativity is the limit of this  
theory
 when $\vert\omega\vert\to \infty$ and $\lambda (\phi)\to 0$; and  
as it is well known, the solar system experiments imply that
$\vert \omega\vert\ge 500.$

The explicit field equations are

\begin{equation}
G_{\mu \nu}=\frac{8\pi T_{\mu \nu}}{\phi}+\lambda (\phi )\quad  
g_{\mu \nu}+
              \omega \phi^{-2}(\phi_{,\mu}\phi_{,\nu}
-{1\over  
2}g_{\mu\nu}\phi_{,\lambda}\phi^{,\lambda})+\phi^{-1}(\phi_{;\mu\nu}-
g_{\mu\nu}\qed\,    \phi),
\end{equation}

\begin{equation}
\qed\,   \phi+{1\over 2}\phi_{,\lambda}\phi^{,\lambda}{d\over d\phi }\ln
\bigl ( {\omega (\phi ) \over \phi}\bigr )+{1\over 2}{\phi \over  
\omega (\phi )}
\bigl [R+2{d\over d\phi }(\phi \lambda (\phi )) \bigr ]
=0,
\end{equation}

\noindent where $G_{\mu \nu}$ is the Einstein tensor. The last  
equation can be substituted by

$$\qed\,    \phi +{2\phi ^2 d\lambda /d \phi -2\phi \lambda (\phi )  
\over 3+2
\omega (\phi )}={1\over 3+2\omega (\phi )}\left (8\pi T-\, {d\omega \over  
d\phi} \phi
_{,\mu} \phi ^{,\mu} \right ), \eqno(2c) $$

\noindent where $T=T_{\mu}^{\mu}$ is the trace of the stress-energy matter tensor.
The divergenceless condition of the stress-energy matter tensor is satisfied if the 
field equation (3) is satisfied (as is shown in  appendix), therefore we shall consider 
equations (2) and (2c) as our field equations.

In what follows we shall assume that
$\omega (\phi)= \omega_0 = constant$, $\lambda (\phi) = {\it c}  
\phi^m$, $\phi=\phi_1 t^q$, with {\it c}, {\it m }and {\it q}  
constants (for recent results when $\omega$ is variable see $^{14}$).
The field equations for this choice of $\omega $ and $\lambda $ and  
with a perfect fluid for the matter content in the isotropic and 
homogeneous line element will be considered,

\begin{equation}
ds^2=dt^2-a^2(t)\left [ {dr^2\over 1-k r^2}+
r^2(d\theta^2+\sin^2\theta d\phi^2) \right ].
\end{equation}

\noindent The field equations are

\begin{equation}
 3(\dot a/a)^2+3k/a^2-{\it c}\phi^m = \frac{8\pi \rho}{\phi}  
+({\omega}/2)(\dot \phi/\phi)^2-
  3(\dot a/a)(\dot \phi/\phi),
\end{equation}

\begin{equation}
 -2({\ddot a}/a)-(\dot a/a)^2-k/a^2+{\it c}\phi^{m}= \frac{8\pi  
p}{\phi}+({\omega}/2)(\dot\phi/\phi)^2+
 {\ddot \phi}/\phi+2(\dot a/a)(\dot \phi/\phi)
\end{equation}

\begin{equation}
 [{\ddot \phi}/\phi+3(\dot a/a)(\dot\phi/\phi)]B=
 2{\it c}(1-m)\phi^m +\frac{8\pi(\rho-3 p)}{\phi},
\end{equation}

\noindent  where B$=3+2\omega$. In the next sections we show some  
exact solutions for these equations when the fluid is a barotropic  
one, $p=\epsilon
 \rho$, including the vacuum case.

\section{The vacuum solutions}

In the case where we neglect all the nongravitational matter we  
have obtained the following solutions.

\subsection{$ k \ne 0$}

\begin{equation}
a= a_1\;t, \;\;a_1=\pm \sqrt{\frac{k m^2}{2+2m+2\omega-m^2}},
\end{equation}

\begin{equation}
\phi=\phi_1\; t^{-2/m}, \;\; c= \frac{2(3+2\omega)}{\phi_1^m m^2},
\end{equation}

\begin{equation}
\lambda=\frac {\lambda_1}{t^2},\;\; \lambda_1=\frac{2(3+2\omega)}{m^2}
\end{equation}

\noindent where $\phi_1$ and $m$ are arbitrary constants . In order to have
real $a_1$, the values of $\omega$ are restricted in the following way,

\begin{eqnarray}
\omega &>& \frac{m^2}{2}-m-1, \; for \; k=1,\nonumber \\
\omega &<& \frac{m^2}{2}-m-1, \; for \; k=-1.
\end{eqnarray}

The value of $\phi_1$ can be related to present day observations  
if we recall that  relation of $\phi$ at the present time $^{1,4}$,

\begin{equation}
\phi_0=G_0^{-1} {4+2\omega\over 3+2\omega},
\end{equation}

\noindent  and the definition of the Hubble constant,

\begin{equation}
H_0=\Bigg( {\dot a\over a} \Bigg)_0={1\over t_0},
\end{equation}

\noindent  (here $t_0$ is the age of the universe), we obtain the  
value of the constant $\phi_1$,

\begin{equation}
\phi_1=\frac{1}{H_0^{2/m} G_0} {4+2\omega\over 3+2\omega}.
\end{equation}

\subsection{$k=0$ solutions}

\begin{equation}
a=a_1 t,
\end{equation}

\begin{equation}
\phi=\phi_1 t^{\frac{-2}{1\pm \sqrt{3+2\omega}}},     
\end{equation}

\begin{equation}
m={1\pm \sqrt{3+2\omega}}, \; c=\frac{2(m-1)^2}{m^2 \phi_1^m},
\end{equation}

\begin{equation}
\lambda=\frac{\lambda_1}{t^2},\;
\lambda_1=\frac{3+2 \omega}
{\omega +2 \pm \sqrt{3+2 \omega}}
\end{equation}

\noindent here $a_1$ and $\phi_1$ are arbitrary constants. The  
value of $\phi_1$ for this case is determined in terms of present  
day values of G and the Hubble parameter, as above, to be

\begin{equation}
\phi_1=\frac{1}{H_0^{2/(1\pm \sqrt{3+2\omega})} G_0}  
{4+2\omega\over 3+2\omega}.
\end{equation}

\section{Barotropic equation of state}

Assuming the equation of state $p=\epsilon \rho$ we have from the conservation 
equation  $\rho=s/{a^{3(\epsilon +1)}}$;  substituting
into the field equations we obtain the following solution

\begin{eqnarray}
a&=&a_1 t,\ \ \phi =\phi_1t^{-(1+3\epsilon)},\nonumber \\
\rho&=&\frac{s}{a^{3(1+\epsilon)}},\nonumber \\
\lambda &=& \frac{\lambda_1}{ t^2}, \, 
\lambda_{1}={{ a_1^2[\omega (1+5\epsilon+3\epsilon^2-
9\epsilon^3 )+2(3\epsilon+1)]+2k(1+3\epsilon)} 
\over  2(1+\epsilon)a_1^2 },
\end{eqnarray}

\noindent where

\begin{eqnarray}
c&=&{{ a_1^2[\omega (1+5\epsilon+3\epsilon^2-9\epsilon^3 )+2(3\epsilon+1)]+2k(1+3\epsilon)} 
\over  2(1+\epsilon)a_1^2 \phi_1^{2 \over (3\epsilon + 1)}}, \nonumber \\
s&=&{ a_1^{1+3\epsilon}\phi_1 \over 8\pi (1+\epsilon) } 
\{2k-a_1^2[\omega (9\epsilon^2+6\epsilon+1)+9\epsilon^2+12\epsilon+1] \}, \nonumber \\
m&=&{2 \over {1 + 3\epsilon}}
\end{eqnarray}

\noindent and

\begin{equation}
\phi_1=\frac{1}{H_0^{(1+3\epsilon)} G_0} {4+2\omega\over 3+2\omega}.
\end{equation}

The above solution is not valid for $\epsilon=-1$, that is for the equation of state that
corresponds to the quantum vacuum. In the next section we consider this particular case.

\section{Vacuum fluid}

Here we want to consider the case when $p=-\rho$, that is the well known 
equation of state for the quantum vacuum. This case means that in  
addition to the contribution of the gravitational theory to the  
cosmological constant we have some other contribution(s) from the  
vacuum spectation value of some quantum fields. Therefore
the effective cosmological constant is $\Lambda_{\rm eff} = 8\pi  
\rho /\phi + c \phi^m$.

For $\epsilon=-1$ we have obtained the following solution,

\begin{equation}
a = a_1 t, \;\; \phi ={ \phi_1}{ t^2}, \;\; \rho = \rho_0=const.,\; \;
\lambda =\frac{ \lambda_1}{t^2}
\end{equation}

\noindent with

\begin{equation}
a_1=\sqrt{\frac{k}{2 \omega-1}}, \;
c=\phi_1(6+4\omega)-8\pi\rho_0
,\;
\lambda_1= 2\frac{[\phi_1(2\omega+3)]-4\pi \rho_0}{\phi_1}, \;\;
m=-1.
\end{equation}

The effective cosmological constant is

\begin{equation}
\Lambda_{\rm eff}= \frac{8\pi\rho }{\phi} + c\phi^m  
=\frac{2(3+2\omega)}{t^2}
\end{equation}

We can see now that in this model regardless of how large is the  
contribution to the cosmological constant of the vacuum energy of
quantum fields the gravitational contribution reduce the value of  
the effective cosmological constant to the present value of $\simeq  
4\omega H_0^2$.

In this case $\phi_1$ can also be related to the
present value of the gravitational constant,

\begin{equation}
\phi_1=\frac{H_0^{2}}{ G_0} {4+2\omega\over 3+2\omega},
\end{equation}

\section{ Final remarks }

In this work we have presented a family of solutions to the general  
scalar-tensor theory of gravity with a potential which plays the  
role of a time dependent cosmological constant.

The time dependence of the cosmological constant in each obtained  
solutions has the form $\lambda \sim t^{-2} $. This dependence occurs 
in a natural way when we use a field potential $\phi \sim t^q $ and we limited 
to ourself to $a \approx t $ that guarantees a coasting period in which the solutions 
are true. This type of solutions in general relativity were studied some time 
ago $^{15,16}$, where the time dependence of $\lambda$ is chosen ad hoc.
The resulting age of the universe, $t_0=1/H_0$ is not in 
conflict with the observational determination$^{17,18, 19}$. Some other 
astrophysical consequences of the models, like nucleosynthesis,$^{20}$ 
remain to be explored.

\section{Appendix}

In this appendix we show that the divergenceless condition is satisfied if the 
field  equation (3) is satisfied.

We start our demonstration with the expression of the  stress-energy  matter tensor, 
which can be written from Eq. (2) as

\begin{equation}
8\pi T^{\mu}_{\nu}= \phi ( R^{\mu}_{\nu}-{1 \over 2}\delta^{\mu}_{\nu}R) - 
{\omega(\phi) \over \phi}(\phi^{,\mu}\phi_{,\nu}-{1\over 2}\delta^{\mu}_{\nu}
\phi_{,\lambda}\phi^{,\lambda})-({\phi_{;}^{\mu}}_{\nu}-\delta^{\mu}_{\nu}\qed\, \phi)
-\lambda(\phi)\delta^{\mu}_{\nu}\phi \, ,
\end{equation}

\noindent taking the divergence of this last equation and rearranging terms, we get

\begin{eqnarray}
0&=&-\phi_{;\mu}{1 \over 2}\delta^{\mu}_{\nu}R-{\omega(\phi) \over 2\phi^{2}}
\phi_{;\mu}\delta^{\mu}_{\nu}\phi_{,\lambda}\phi^{,\lambda}+{d\omega(\phi) \over 
d\phi}{1 \over 2\phi}\phi_{;\mu}\delta^{\mu}_{\nu}\phi_{,\lambda}\phi^{,\lambda}
-{\omega(\phi)  \over \phi}{\phi^{;\mu}}_{;\mu}\phi_{,\nu} \nonumber \\
& &+{\omega(\phi) \over \phi}\phi^{,\mu}\phi_{,\nu;\mu} 
-\phi {d\lambda(\phi) \over d\phi}\phi_{;\nu}-\lambda(\phi)\phi_{;\nu}+\phi_{;\mu}
R^{\mu}_{\nu} 
+{\omega(\phi) \over \phi^{2}}\phi_{;\mu}\phi^{;\mu}\phi_{,\nu} \nonumber \\
& &-{d\omega(\phi) \over d\phi}{1\over \phi}\phi_{;\mu}\phi^{,\mu}\phi_{,\nu} 
+{\omega(\phi) \over \phi}\phi_{,\lambda,\nu}\phi^{,\lambda}
-{\phi_{;}^{\mu}}_{\nu;\mu}+(\qed\, \phi)_{;\nu} \, ,
\end{eqnarray}

\noindent  simplifying this last equation:

\begin{eqnarray}
0&=&-{1\over 2}\phi_{;\nu}R+{1\over 2}{\omega(\phi) \over \phi^{2}}\phi_{;\nu}
\phi_{,\lambda}\phi^{,\lambda}-{d\omega(\phi) \over d\phi}{1 \over 2\phi}
\phi_{;\nu}\phi_{,\lambda}\phi^{,\lambda}-{\omega(\phi) \over \phi} 
{\phi^{,\mu}}_{;\mu}\phi_{,\nu} \nonumber \\
& &-\phi{d\lambda(\phi) \over d\phi}\phi_{;\nu} 
-\lambda(\phi)\phi_{;\nu}+\phi_{;\mu}R^{\mu}_{\nu}-{{\phi_{;}^{\mu}}_{;\nu}}_{;\mu}+
(\qed\, \phi)_{;\nu} \, .
\end{eqnarray}

Taking into account the identities $\phi_{;\mu}R^{\mu}_{\nu}=\qed\, (\phi_{;\nu})-
 (\qed\, \phi )_{;\nu}$, in the previous equation, we have

\begin{equation}
0=-{\omega(\phi) \over \phi}\phi_{;\nu}\Bigg[ \qed\, \phi +{1\over 2} \phi_{,\lambda}
\phi^{,\lambda}{d \over d\phi} ln \Bigg( {\omega(\phi) \over \phi}\Bigg) +
{1\over 2}{\phi \over \omega(\phi)}\bigg[ R+2{d \over d\phi}(\phi \lambda(\phi)) \bigg]
\Bigg]
\end{equation}

This is the field equation (3) times the factor $-{\omega(\phi) \over \phi}\phi_{;\nu}$, 
therefore if this field equation is satisfied, then the divergenceless condition is satisfied 
too and it is enough to use  equations (2) and (2c) as  the field equations of our 
system from which we obtain Equations (5), (6) and (7).

\section{ Acknowledgment}
\bigskip
\bigskip
This work was partially supported by CONACyT GRANT 1861-E9212.

\newpage

\nonumsection{References}
\noindent
%References are to be listed in the order cited in the text. %Use
%the style shown in the following examples. For journal %names,
%use the standard abbreviations. Typeset references in 9 pt %Times
%Roman.

\end{document}